# Implementation of Sensor Network using Efficient CAN Interface

[1] **Yeshwant Deodhe,** [2] **Swapnil Jain,** [3] **Ravindra Gimonkar**

[1] Deptt. of ElectronicsYCCE,Nagpur

[2] Department of Electronics and Communication, DMIETR, Wardha

[3] Deptt. of Electrical Engg. YCCE, Nagpur

**Abstract -** Sensors monitored by centralized system, that may be used for controlling and monitoring industrial parameters (Temp, Pressure, Speed, Torque) by using CAN interface. In this paper we presents a comprehensive overview of controller area networks, their architecture, protocol, and standards. Also, this paper gives an overview of CAN applications, in both the industrial and non-industrial fields. Due to CAN reliability, efficiency and robustness, we also propose the extension of CAN applications to sensor network. In this paper, a framework of sensor network for monitoring industrial parameters is explained where sensors are physically distributed and CAN is used to exchange system information. CAN (Controller Area Network) is a high integrity serial bus protocol that is designed to operate at high speeds ranging from 20kbit/s to 1Mbit/s which provide an efficient, reliable and very economical link.

## 1. Introduction

Major communication networks can be divided into four types, namely: IP Core Network/Internet, Wireless LAN, 3G/4G Cellular Network[1]. The common usage for these networks is to carry text, audio and video content. Recently, some of these networks have been utilized in industrial automation to monitor and control industrial plants. Another type of networks is the Controller Area Network. CAN is intended as a communication network between the control units in vehicles. Nowadays, CAN applications are gaining ground and it is extending to industrial automation including marine and aircraft electronics, factories, cars, trucks and many others. The backbone of the Controller Area Network is a fast serial bus that is designed to provide a reliable, efficient and a very economical link between sensors and actuators. CAN use a twisted-pair cable to communicate at speeds up to 1 Mbits/sec, with up to 40 devices. CAN was originally developed to simplify the wiring in automobiles. In the past, automobile manufacturers used to connect devices in vehicles using point-to-point wiring systems. As more electronics and controllers are deployed to monitor and control vehicles, wiring started to become more complex, bulky, heavy and expensive. Automotive industry starts to reduce massive wires complexity with dedicated CAN link that provides low-cost, robust network, and multi-master communication system. Figure 1 shows the efficiency and the wiring-reduction caused by implementing CAN among multiple devices. Thus causing wiring-reduction, cost reduction with CAN.

## 2. Hardware

Three major elements can be found in the sensor network prototype: a sensor node module, a computer server module and a computer client module. Each element is discussed below. a) Sensor node module: hardware block diagram of the sensor node module is shown in Fig. 2. Each module consists of a free scale 8 bit AVR processor with built-in CAN I/O controller, an SPI connected analog to digital circuit, which digitizes the signals from a series of temperature sensors. All sensor nodes are interconnected to a differential two-wire physical CAN bus (CAN High and CAN Low), which is controlled by a central PC computer.

CAN was originally developed by Robert Bosch (Germany, 1986) when Mercedes requested a communication system between three electronic control units in vehicles. Point to point communication was not suitable anymore, and the need of using a multi-master communication system became imperative. Although this origin can be traced to automotive industry, industrial automation rapidly showed the need of using such a popular bus system.[2-3] CAN port is a two-wire, half duplex, high-speed network system that can reach a throughput up to 1 Mbits/sec. transmitted. The transfer layer defines the Kernel of the CAN protocol. It is responsible for presenting and accepting the received/transmitted messages to/from the upper layer.

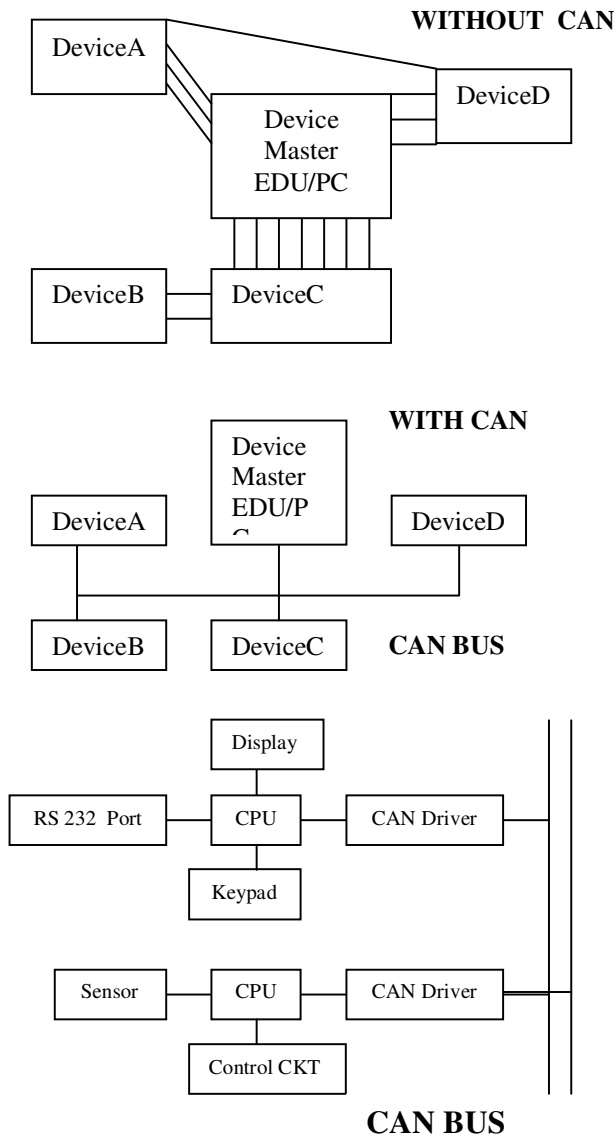

Fig 1 Reduction in wiring with CAN

Fig 2 A hardware block diagram of the sensor node module

## 3. CAN Architecture, Standards and Protocol

Data, control commands and devices status can be transmitted and/or received in well structured frames. Theoretically, CAN is capable of linking up to 2032 devices on a single network; due to hardware limitation, only 110 nodes can be linked-up to construct a single network.[2] Similar to the traditional OSI model and IP model CAN have a 4-layer protocol: physical layer, transfer layer, object layer, and application layer as shown in figure 3.[3]. Fig.3. Layered Architecture of a CAN node .The physical layer defines how the signals are These messages are sent and received using the following properties: bit timing, message framing and arbitration, acknowledgment, error detection and signaling, in addition to fault confinement. The object layer is responsible for message filtering and handling. The object layer and the transfer layer are both combined as the data link layer defined by the ISO/OSI model. The application layer is the upper layer, through which the user interferes. CAN (Version 2.0) has two different standards: CAN 2.0 A, standard CAN, using 11 bits for node identification; and the other standard is CAN 2.0 B or extended CAN, using 29 bits for node identification. With 11 bits, 2,048 unique messages are possible, whereas 536 million unique messages are possible with 29-bit identifier In a controller area network, all nodes - recipients - can see all messages and can accept or ignore any messages according to the system design. There are two ISO standards classifying the two CAN's physical layer in terms of data rate; ISO 11898 which can handle speed up to 1 Mbit /sec and ISO 11519 that can handle speed up to 125 Kbit /sec, as shown in table 1.

Fig 3. Four layer protocol of CAN

## 4. System Summarization

First of all, it should be defined function of each node, in order to sure the number, type, signal characters of node control quantity. This is the first step of proceeding the control system to realize networking. Secondly, it should be chosen node controller and relevant CAN components.

For the functions of every node are single comparatively, and the quantities of data are also small, the demand of CPU is reduced greatly, which could be fulfilled by AVR processor. The parts of CAN bus are mainly constructed by controller port, bus transceiver and parts of I /O. Finally, according to agreement of physical layer in CAN bus to choose the medium of bus, design the wiring project, and connect it to be control network.

### 4.1. Capability Comparison between RS-232 and CAN Bus

Industry facilities communication often related to a lot of hardware and soft ware product. It is used to connect the protocol between standard computer flat and industry automatic application facilities. Moreover, the facilities and protocols in which were used are various. Therefore, it is hoped that most automatic application facilities can perform simple serial orders, especially hoped that these orders are compatible with the standard serial ports in PC or accessional serial port boards. RS-232 which has widest application in PC and communication industry is one of the serial ports. RS-232 is defined as a sort of single end standard which could increase the communication distance in serial communication with low-velocity. With the communal signal ground between sending port and sink of RS-232, it cannot use signal with two ports. Otherwise, the common mode noise will be coupling into signal system.

CAN is one of the field bus which is widest application internationally full named "Controller Area Network". As a sort of serial communication bus with multi-mainframe mode, the basic design criterion of CAN demands high-velocity and better capability of contradicting electromagnetism disturb, even demands to check any mistakes which are produced in communication bus. When the distance of signal communication reached 10km, CAN still provide digital communication velocity with 50kbit/s [5].

### 4.2 System Composition of CAN and RS-232 Convertor

It is used singlechip AT89C51 as microprocessor, SJA1000 as microcontroller of CAN in designing the transition equipment which change RS-232 into CAN. As shown in Fig. 1. SJA1000 could process the frame in communication data for integrating the function in physical layer and data link layer of CAN protocol [5]. As an interface between CAN controller and physical bus, PCA82C250 is used to provide the differential sending of bus and the differential receiving of CAN controller. There are three different working could be chosen through the Pin3 of PCA82C250 (high speed, slopecontrol, readiness). When Pin3 connects to earth, it is working in high speed.

Max is used to complete the level transition from RS-232 to interface chip in micro-controller.

### 4.3 Circuit Design Of System Hardware

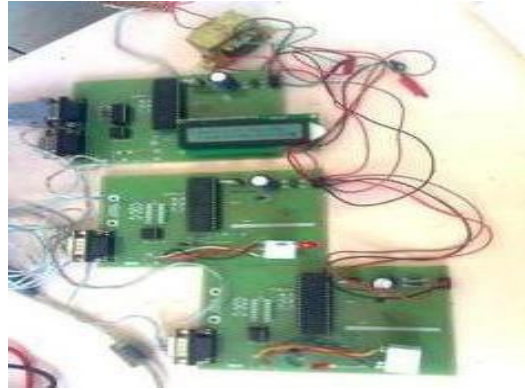

Fig 4. Hardware circuit design

Hardware circuit of system is mainly constructed by level transition circuit of RS-232, controller and transceiver of CAN. The collective circuit PCB of hardware design is followed as Fig. 2.

*A. Detailed Design of Every Part* As an unattached level transition controller, this brainpower transducer relates to fetch temperature, brainpower switch in the mode of sending and receival, setting of communication mode, level transition of RS-232 and some other aspects. Material description shown as:

### 4.3.1 Level Transition Circuit of RS-232

In definition of normal RS-232 interface, TXD, TRS and DTR are the level output of RS-232. In the system facilities of computer data collection and industry, RS-232 interface is the most familiar communication standard. It is prescribed the meanings of ambipolar level data "0" and "1" were expressed together by voltage amplitude and level polarity in RS-232 standard interface. Through fair and foul, the maximum value is ±15v; it is defined 4 logical levels in RS-232 standard interface. For input, it is prescribed +3v~+15v as data "0" and control line level in connection state, while -3v~-15v as data "1" and control line level in disconnection state. When level absolute value under 3v, it is uncertain state. For fan-out, it is prescribed +5v~+15v as data "0" and control line level in connection state, while -5v~-15v as data "1" and control line level in disconnection state. When level absolute value under 5v, it is uncertain state. The prescriptive logical level of RS-232 was different from the current microprocessors and single chip. Therefore, it is should be converted the level between microprocessor and RS-232 in actual application. This conversion should be completed by Max232 in design.

4.3.2 Receival and Sending

*P*rocessor AVR is a core of the module, which could complete the function of application layer in CAN bus. It is chosen AVR as main controller with which SRAM 1 K bytes, FLASH ROM 16K byte in system self programmable,512 bytes, external & internal interrupt sources, 32 programmable I/O lines. For the message queue quoted in program, it is extended static state RAM with 1k, CAN controller with Basic work module. It is chosen SN651051 as CAN transceiver which could connected CAN controller to CAN physical network belongs to the interface of control circuit and physical transfers' circuit. MAX232 is an apparatus for conversion between TTL/COMS level and EIA-232 level which could connect singchip's UART and microcomputer's COM to realize the full duplex communication of the both.

4.3.3 Realization of Node Circuit in CAN Bus

Node message in network could be divided into different PRI to fulfill different real time demanding. CAN with multi-mainframe mode, any nodes in the network could send message to other nodes forwardly in any time, without the difference of principal and subordinate node.

When many nodes send message to bus simultaneously, CAN choose the non destructive bus arbitration technique. Node with low priority will be sent in turn, while node with high priority will be sent normally without any influences to save the time of conflict arbitrage. In time of sending, the losing arbitrage or the frame which go to pot by making mistakes will be resent automatically. The direct communication distance of CAN will be 10km (velocity under 5kbps), communication velocity will maximize 1 Mbps (here, communication distance maximize 40m). Message of every frame in CAN all have CRC checkout and other debug measures to insure low error rate. In the condition of serious errors, CAN nodes have function of closing and existing bus automatically to make other nodes without any influences?

*B.* Problems Should Be Settled in Hardware Design Some problems would be appeared in design of hardware circuit. It is shown as follow.

*1)* Connection circuit between MAX232 and serial-port*:* Fig.5 is the connection circuit between MAX232 and computer serial-port. The second pin of serial-port is sink which connects to 14 or 9 pin of MAX, the third pin is sending terminal which connects to 13 or 8 pin of MAX. The data terminal of wireless sending module connects to 10 or 11 pin.

*2)* Connection of AVR receiver:
Temperature sensors are connected to PA0,PA1 inputs of AVR ,which is converted to its digital equivalent through inbuilt 10 bit ADC ,RX0 and TX0 of AVR Processor connected to SN 651050.

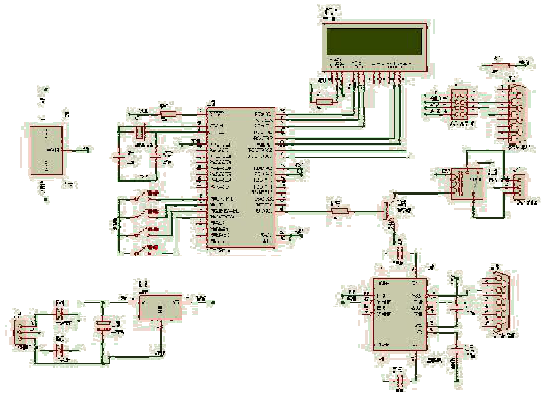

Fig. 5 Connection circuit between MAX232 and computer serial-port

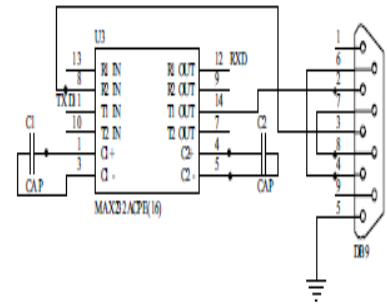

Fig. 6 Connection from RS-232 to MAX232

SN 651050 as CAN bus transceiver is an interface apparatus between CAN controller and CAN bus which send it with different mode for CAN. Four switches are connected to PB0 to PB3 input pins of AVR. Depending upon status of 4 switches, the required temperature of the system is monitored there is only used the first data channel in the design pin no 14 is RxD,pin no 15 is TxD of AVR used to communicate data It is shown as Fig. 7.

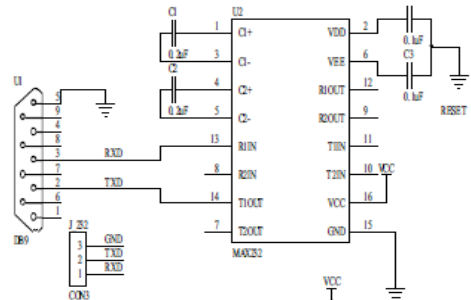

**Fig 7** Circuit diagram for sensor network The experimental results demonstrate the efficiency and accuracy of the applied method. Table 1 lists acquired data

Table 1 acquired data for 1 hour

| Test no | Set temperature in degree | Measured temperature in degree |
|---|---|---|
| 01 | 20.00 | 20.01 |
| 02 | 22.50 | 22.50 |
| 03 | 23.00 | 22.99 |
| 04 | 25.30 | 25.30 |
| 05 | 30.00 | 30.01 |
| 06 | 16.00 | 16.01 |
| 07 | 19.50 | 19.49 |
| 08 | 22.00 | 22.01 |
| Sum | 178.30 | 178.32 |

We can see that temperature measuring error is 0.02 degree centigrade and the temperature accuracy is 0.02 % relative to the temperature scale of 100 degree centigrade

## 5. Conclusion

These sensor networks referred in the paper are just used some cheap and universal components, like AVR, MAX232, SN 651050 etc. For circuit design having the advantages of briefness and integrity, it has the characters of simple configuration and low-cost which compared to some similar industrial products.

There are some excellencies of choosing CAN bus technique: (1) Speedy response, better anti-jamming capability; (2)lowest malfunction rate, high communication baud rate, great data transfer with network which was constructed by two twisted-pairs; (3) simple configuration, great scalability, high security and convenient setting, servicing [6]. The system based on field bus CAN has a very nicer application foreground in control system domain for a good many excellences of CAN.

Switched RS-232communication network to CAN communication network could realize to construct network with multi drop of RS-232 and long-distance-communication conveniently. Moreover, it could substitute the CAN interface card with costly price in current market to realize data communication between PC serial interfaces and CAN bus fleetly and exactly. The configuration of hardware debugged and passed, which have nicer transplant and extensibility.

## References

[1] Minu A Pillai "Implementation of Sensor Network for Indoor Air Quality Monitoring using CAN Interface" International Conference on Advances in Computer Engineering-2010.
[2] Mario Alves, Miguel Pereira, Helena Geirinhas Ramos, "CAN protocol: A laboratory prototype for Field Bus Applications", *XIX IMEKO World Congress, Fundamental and Applied Metrology*, pp. 454-457, September 2009.
[3] R.P.Deshmukh " Design and implimentation of sensor network using CAN bus" vol-2 International conference on computer applications 2012.
[3] GAO Jun, "Design of the Intelligent Measuring & Controlling System Based on CAN Field Bus", *Light Industry Machinery*, Vol.24, No. 2,2006, pp.103~106.
[4] Kyung Chang Lee and Hong-Hee Lee, "Network-based fire-Detection System via Controller Area Network for Smart Home Automation", *IEEE Trans. Consum. Electron*, 1094 vol. 50, no. 4, pp. 1093-1100,July 2004.
[5] J.V. Capella, J.J. Serrano, J.C. Campelo, A. Bonastre, R. Ors and P.Bradbury, "A new Control System for Citric Fruit Conservation and Maturation based on CAN and Internet networks", *CAN inAutomation, ICC 2003*, pp. 6-14 to 6-21.
[6] Th. Zahariadis,"Evolution of the Wireless PAN and LAN standards", Computer Standards & Interfaces, Vol. 26, No. 3, May 2004, Pp 175-185.
[7] "Controller Area Network – CAN Information," http://www.algonet.se/-staffanun/developer/CAN.htm, 3 November 2005.
[8] Bosch, R. "CAN SPECIFICATION (Version 2.0),"Germany: Stuttgart, 1991

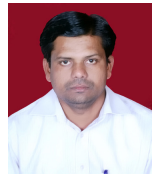

Prof. Yeshwant A.Deodhe ,Assist. Professor, Deptt. Of Electronics, YCCE, Nagpur  has competed B.E. Electronics in 1996 from Nagpur university . M.Tech Electronics in 2011 from Nagpur university. Five Research publications in IEEE international conferences  in India. Area of specialization is VLSI and communication engineering.

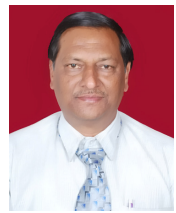

Prof.Ravindra M. Gimonkar,Assist.professor,Deptt. Of Electrical engineering,,YCCE,Nagpur
Qualification :-- M-Tech(Integreted power system)
two papers published in national conferences on the topic  of "reliability analysis of hvdc link." Area of specialization is electrical power systems.